# Thickness determination of $MoS_2$, $MoSe_2$, $WS_2$ and $WSe_2$ on transparent stamps used for deterministic transfer of 2D materials

Najme S. Taghavi[1,2], Patricia Gant[1](✉), Peng Huang[1,3], Iris Niehues[4], Robert Schmidt[4], Steffen Michaelis de Vasconcellos[4], Rudolf Bratschitsch[4], Mar García-Hernández[1], Riccardo Frisenda[1](✉), Andres Castellanos-Gomez[1](✉)

[1] Materials Science Factory, Instituto de Ciencia de Materiales de Madrid (ICMM), Consejo Superior de Investigaciones Científicas (CSIC), Sor Juana Inés de la Cruz 3, 28049 Madrid, Spain.
[2] Faculty of Physics, Khaje Nasir Toosi University of Technology (KNTU), Tehrān 19697 64499, Iran.
[3] State Key Laboratory of Tribology, Tsinghua University, Beijing 100084, China.
[4] Institute of Physics and Center for Nanotechnology, University of Münster, 48149 Münster, Germany.
————————————————

Address correspondence to Patricia Gant (patricia.gant@csic.es), Riccardo Frisenda (riccardo.frisenda@csic.es) and Andres Castellanos-Gomez (andres.castellanos@csic.es)

**ABSTRACT**

Here, we propose a method to determine the thickness of the most common transition metal dichalcogenides (TMDCs) placed on the surface of transparent stamps, used for the deterministic placement of two-dimensional materials, by analyzing the red, green and blue channels of transmission-mode optical microscopy images of the samples. In particular, the blue channel transmittance shows a large and monotonic thickness dependence, making it a very convenient probe of the flake thickness. The method proved to be robust given the small flake-to-flake variation and the insensitivity to doping changes of $MoS_2$. We also tested the method for $MoSe_2$, $WS_2$ and $WSe_2$. These results provide a reference guide to identify the number of layers of this family of materials on transparent substrates only using optical microscopy.

**MANUSCRIPT TEXT.**

Since the isolation of graphene in 2004[1], the mechanical exfoliation method (also called Scotch tape method) has established itself as one of the most used techniques to produce 2D materials [2–5]. Its facile implementation combined with the high quality of the produced samples are most likely the reasons behind the success of this technique. Mechanical exfoliation, nonetheless, yields randomly distributed flakes of various thicknesses and sizes on the surface of the substrate. This limitation has been overcome to a great extent through the development of rapid methods to find and identify thin flakes based on the observation of the apparent color when they are transferred onto a $SiO_2$/Si surface [6–12]. On this substrate, there is a dependency of the apparent color of the flake with its thickness due to thin-film interference effects and many epi-illumination (reflection mode) microscopy-based methods have been developed to identify 2D materials and to determine their number of layers [6–12]. An alternative way to overcome the limitations of mechanical exfoliation relies on the use of experimental tools that allow for the deterministic placement of flakes onto any desired sample position with micrometer accuracy [13–19]. These transfer techniques opened the field of van der Waals heterostructures [18,20–23]. To carry out the deterministic placement, the isolated flakes have to be deposited onto the surface of a polymer-based stamp that is used as a release layer or sacrificial carrier substrate. These stamps are made of transparent material to allow for the accurate alignment of the flake to be transferred with the acceptor surface by inspection with an optical microscope. Unfortunately, on the surface of the transparent stamps most of the previously developed optical identification methods, based on the apparent colors, are less straightforward to implement and less effective and other complementary techniques such as Raman or photoluminescence (which are slower and more complex than simple optical microscopy) are typically used to determine the number of layers of the flakes deposited on the polymeric stamps [24–27]. Due to the increasing interest on the use of deterministic placement methods to assemble nanodevices and van der Waals heterostructures, the development of alternative methods to determine the thickness of 2D materials on the surface of the transparent polymeric stamps is extremely relevant for the 2D community. That is exactly the goal of this work, to provide a fast and reliable method to identify TMDC flakes on transparent polymeric stamps and to determine their number of layers.

Here, we demonstrate that the quantitative analysis of transmission optical mode images of TMDC flakes on the surface of transparent polydimethylsiloxane (PDMS) stamps is a reliable method to accurately determine their thickness. We compare the results of the quantitative analysis of the transmission mode optical images with results obtained via micro-reflectance spectroscopy [28], photoluminescence and Raman spectroscopy to verify its reliability. In order to test the limitations of this method, we probe its sensitivity concerning the doping level of the sample by measuring $MoS_2$ samples with different intentional doping. We found that the determination of the thickness with the analysis of transmission mode images is rather independent on the doping level, giving similar results for all the studied samples. Finally, we extended the study to other TMDCs to provide a reference guide to identify the number of layers of this family of materials. As the measurement of the transmittance is a differential measurement, the effect of the substrate is accounted for and thus these results could be extrapolated to other transparent stamp surfaces like poly(methyl methacrylate) (PMMA) or polycarbonate/hexagonal boron-nitride (PC/hBN).

Figure 1a shows a transmission mode optical microscopy image of a $MoS_2$ flake transferred onto a Gel-Film substrate (a commercially available polydimethylsiloxane film, by Gel-Pak) which is typically used as a stamp for deterministic dry transfer of 2D materials [16,29]. Figure 1b displays line profiles of the intensity of the red, green and blue channels extracted from the image in Figure 1a. The transmittance of the different channels is calculated by normalizing the red, green and blue (RGB) channel images to the intensity of these channels measured on the bare substrate. From the line profiles, one can see that the transmittance of each channel changes stepwise with the number of layers. To determine the feasibility of using the transmittance to determine the number of layers one needs to characterize the statistical variations of the transmittance in different $MoS_2$ flakes and the uncertainty associated to these measurements. Therefore, we acquired transmission mode optical microscopy images of 202 $MoS_2$ flakes and we compiled three histograms by binning the measured transmittance of the flakes for each color channel. Figure 1c shows the histograms of the red, green and blue channels plotted versus 1-transmittance. Although all the histograms display prominent peaks that correspond to different number of layers, the histogram of the blue channel shows a larger separation between the peaks (while maintaining a similar peak width) making it easier to distinguish between 1, 2, 3 or 4 layers of $MoS_2$. The higher contrast of the blue channel is expected as $MoS_2$ (and $MoSe_2$, $WS_2$ and $WSe_2$ as well) present a strong excitonic feature (sometimes referred to as C exciton in the literature) in the blue part of the spectrum that yields a strong optical absorption in that wavelength range.[30,31]

In order to verify the reliability of this method to determine the number of layers we benchmarked it with other commonly used spectroscopic techniques. In Figure 2a the experimental data employed to build up the histogram in Figure 1c is displayed in a scatter plot. We have selected 4 flakes with different thickness in the range between 1 layer and 4 layers, respectively labeled as 1L, 2L, 3L and 4L accordingly to their blue channel transmission, and we have carried out micro-reflection, Raman and photoluminescence measurements to have three independent methods to determine the thickness. Figure 2b shows the differential reflectance spectra acquired on these flakes where the intensity and the energy of the A exciton monotonically changes from 1L to 4L. The intensities and exciton positions obtained in the spectra agree with the values expected for 1L to 4L,[31] which means a correct assignment of the thickness estimated by the blue channel values [32]. Figure 2c compares the Raman spectra measured on the same flakes (vertically displaced by 0.1 for clarity) with a 532 nm excitation laser. We verify the number of layers of the samples from the difference between the $A_{1g}$ and $E^1_{2g}$ phonon [33,34] energies in Raman and check again the accuracy in the determination of the number of layers using the blue channel transmittance. Figure 2d shows photoluminescence measurements for 1L, 2L, 3L and 4L flakes, also measured with a 532 nm excitation laser. The intensity and the position of the A exciton dramatically depends on the number of layers and it confirms that the assignment of number of layers determined by the blue channel transmission is reliable.[35,36]

We further test the analysis of the blue channel transmission images by studying $MoS_2$ samples synthesized with intentional substitutional metal atoms at the Mo sites to enquiry about the robustness of this technique against a moderate variation of the chemical composition that lead to a big change in the electronic properties.[37,38] We follow the synthesis method described in references [37–40] for the growth of Fe-doped, Ni-doped, Nb-doped and Co-doped $MoS_2$ crystals. In all cases a 0.5% of dopant material has been added in the ampoule for the synthesis of these doped $MoS_2$ samples which lead to a final doping level in the 0.3-0.4% range. Figure 3 compares the histograms of the blue channel transmittance constructed by analyzing 50 flakes of each doped $MoS_2$ material. This comparison clearly illustrates how our method is rather insensitive to variations of the chemical composition (that induces strong changes in the doping level of the material). We note that other optical spectroscopic methods to determine the thickness of transition metal dichalcogenides such as Raman and photoluminescence spectroscopy are strongly dependent on the doping level of the samples and on slight variations of the chemical composition.[41–45]

We extended this method to other 2D materials of the TMDCs family to provide a general guide to identify these materials through the analysis of the blue channel of transmission images. Figure 4 compares the blue channel transmission histograms built up after analyzing 202 $MoS_2$ flakes, 200 $MoSe_2$ flakes, 200 $WS_2$ flakes and 200 $WSe_2$ flakes. In all the cases the histograms show well-resolved peaks indicating that one can unambiguously determine the number of layers through the quantitative analysis of the blue channel of the transmission mode optical images.

In summary, we have introduced a very simple and fast yet reliable method to determine the number of layers of $MoS_2$, $MoSe_2$, $WS_2$ and $WSe_2$ deposited on a PDMS stamp, used for deterministic placement of 2D materials. Moreover, being a differential measurement, the effect of the substrate is removed and it could be extended to other transparent substrates. We have demonstrated that the transmittance, extracted from the blue channel of transmission mode optical images, monotonically depends on the number of layers. We have benchmarked the layer assignment done with this method with other extended spectroscopic techniques (micro-reflection, Raman, photoluminescence) finding an excellent agreement. Interestingly, this method is robust enough to provide an accurate layer determination even for samples with different doping level. In view of all this, the quantitative analysis of the transmission mode images can become a powerful method to select the flakes of 2D materials deposited onto transparent polymeric stamps prior to their deterministic transfer.

**Experimental**

*Materials:*

$MoS_2$ samples were prepared out of a bulk natural molybdenite crystal (Moly Hill mine, Quebec, Canada). $MoSe_2$ and $WSe_2$ samples were prepared out of bulk synthetic crystals grown by physical vapour transport method (provided by Prof. Rudolf Bratschitsch). $WS_2$ samples were prepared out of a bulk synthetic crystal grown by physical vapour transport method at Tennessee Crystal Center. The Fe-doped, Ni-doped, Nb-doped and Co-doped $MoS_2$ crystals were grown following the protocols described in Refs. [37–40]. A 0.5% of dopant material has been added during the synthesis of these doped $MoS_2$ samples which leads to a final doping level in the 0.3-0.4% range.

The transparent stamp substrate used in this work is a commercially available polydimethylsiloxane-based substrate manufactured by Gel-Pak® (Gel-Film® WF X4 6.0 mil). The TMDC flakes are exfoliated from the bulk crystals and transferred onto the surface of the Gel-Film stamp with a Nitto 224SPV adhesive tape.

*Optical microscopy:*

Optical microscopy images have been acquired with three different upright metallurgical microscopes, an AM Scope, a Motic BA MET310-T and a Nikon Eclipse CI, obtaining identical results. The transmission mode images have been acquired with an achromat condenser lens (N.A. 0.85) that ensures Köhler illumination, yielding to a homogeneous illumination spot of approximately 4 $mm^2$ on the sample. The light was collected with a 50x magnification plan achromatic objective (NA 0.55). Two different digital cameras were tested during this work, an AM Scope mu1803 camera with 18 megapixels and a Canon EOS 1200D, also providing identical results.

*Image analysis*:

The quantitative analysis of the transmittance of the flakes and the rippling wavelength has been carried out using Gwyddion® and ImageJ software.[46,47]

**Acknowledgements**

We thank Prof. Der-Yuh Lin and Prof. Tsung-Shine Ko for providing the doped $MoS_2$ samples. NT acknowledges to the Ministry of Science, Research and Technology of Iran. ACG and PG acknowledge funding from the European Commission Graphene Flagship (Grant Graphene Core 2 785219). RF acknowledges support from the Netherlands Organization for Scientific Research (NWO) through the research program Rubicon with project number 680-50-1515. ACG acknowledge funding from the European Research Council (ERC) under the European Union's Horizon 2020 research and innovation programme (grant agreement n° 755655, ERC-StG 2017 project 2D-TOPSENSE). D.-Y.L. acknowledges the financial support from the Ministry of Science and Technology of Taiwan, Republic of China under contract No. MOST 107-2112-M-018-002..

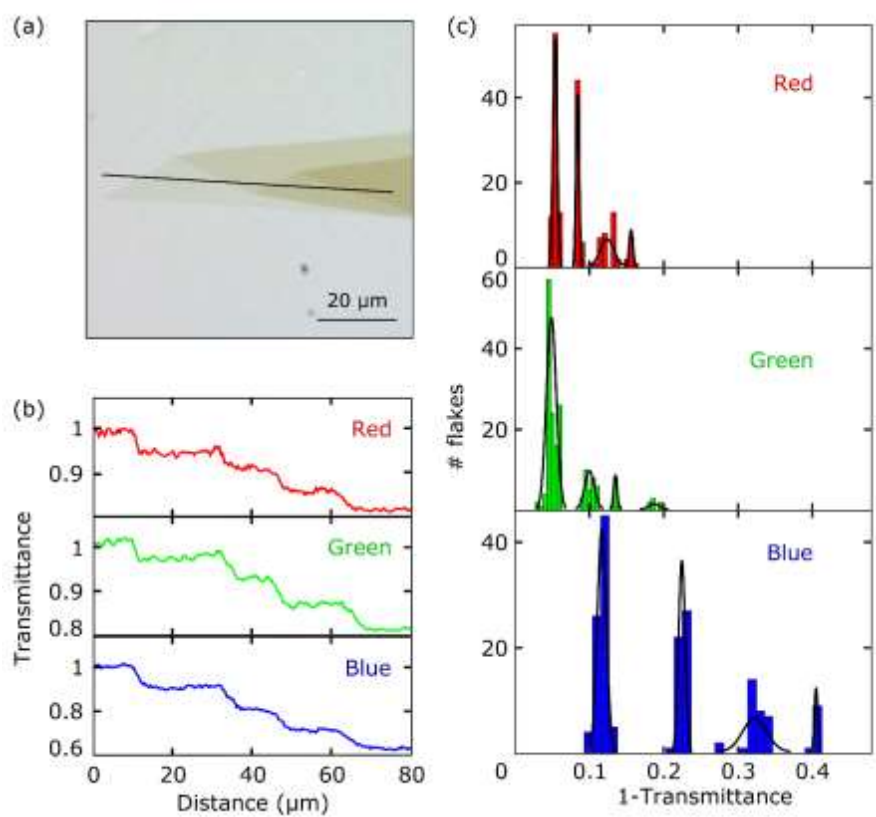

**Figure 1.** (a) Optical image in transmission mode of a MoS$_2$ flake with different thicknesses. (b) Line profiles of the intensities of the red, green and blue channels of the image in 1a.   (c) Histograms of the 1-transmission value in the RGB channels for several MoS$_2$ flakes.

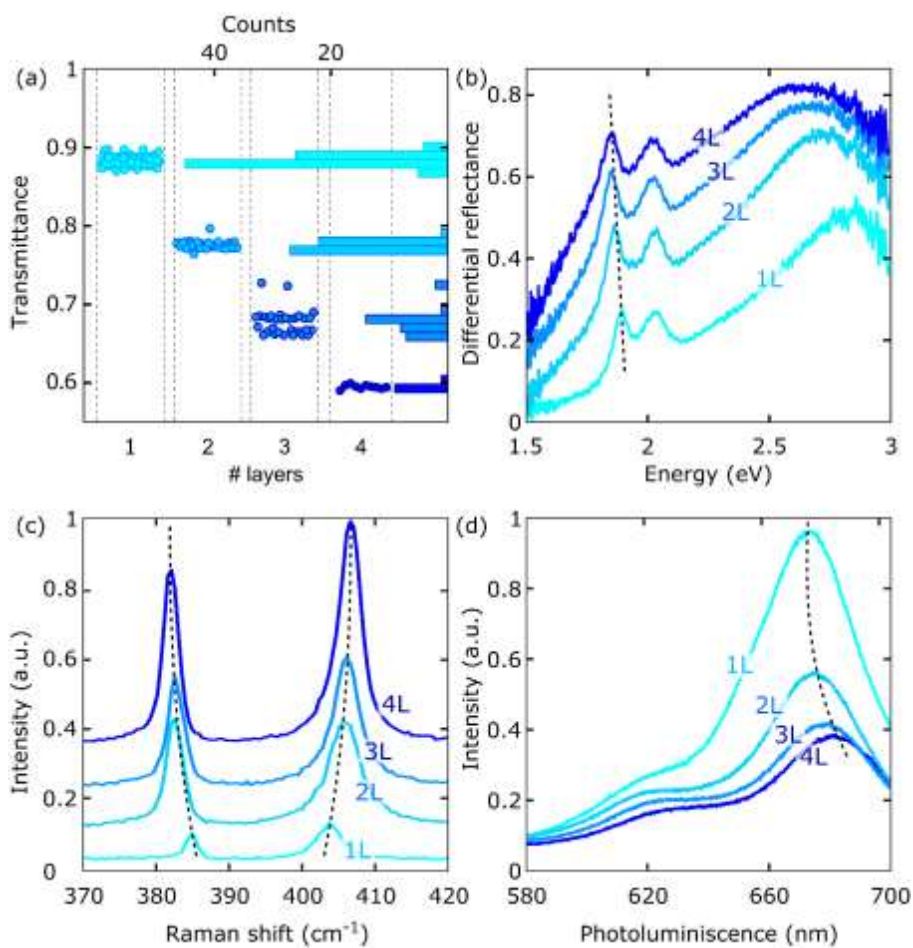

**Figure 2.** (a) Scatter plot and histogram of the blue channel transmission values for several $MoS_2$ flakes. (b) Differential reflectance spectra for $MoS_2$ with different number of layers. (c) Raman spectra for $MoS_2$ with different number of layers. (d) Photoluminescence of $MoS_2$ for different number of layers.

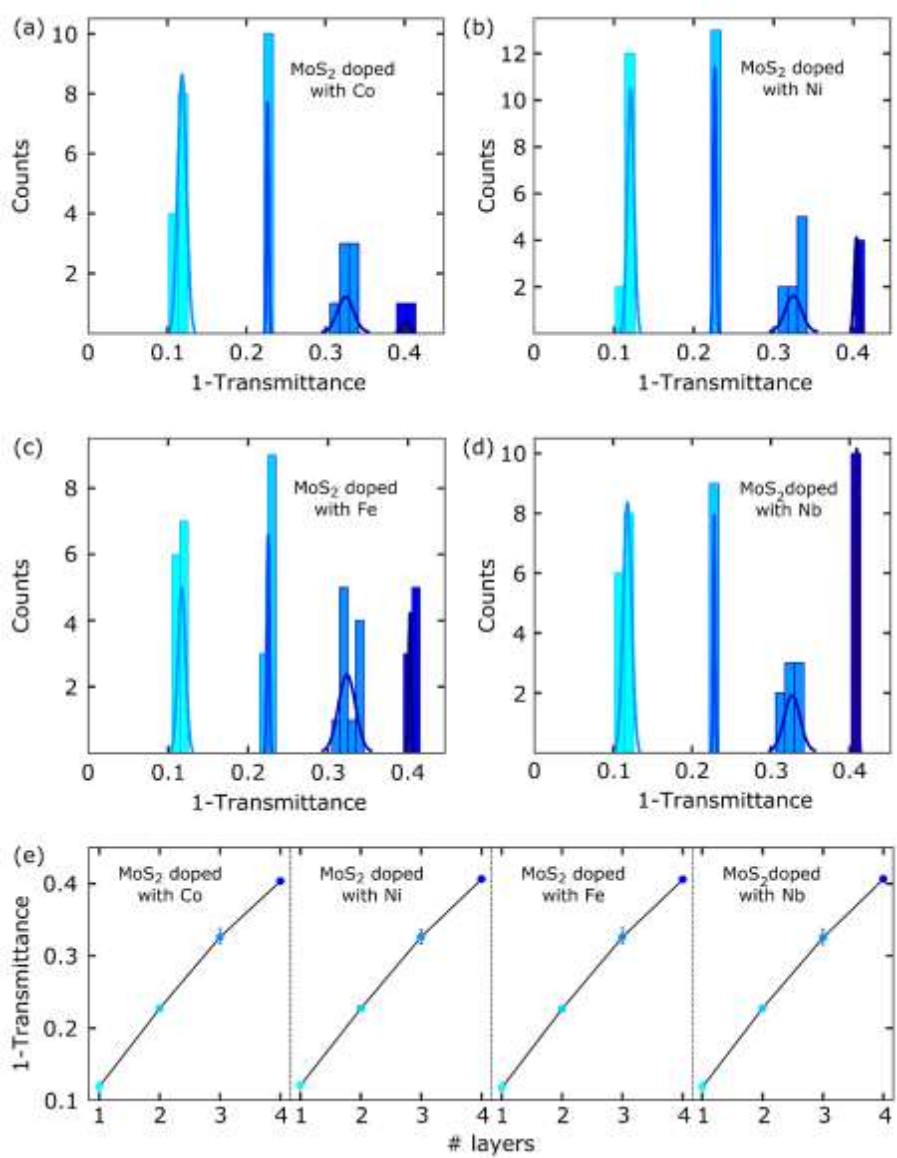

**Figure 3.** Histograms of the 1-transmission value in the blue channel for $MoS_2$ flakes doped with (a) Co, (b) Ni, (c) Fe and (d) Nb. (e) Average of the 1-transmission values for 1L to 4L in each material.

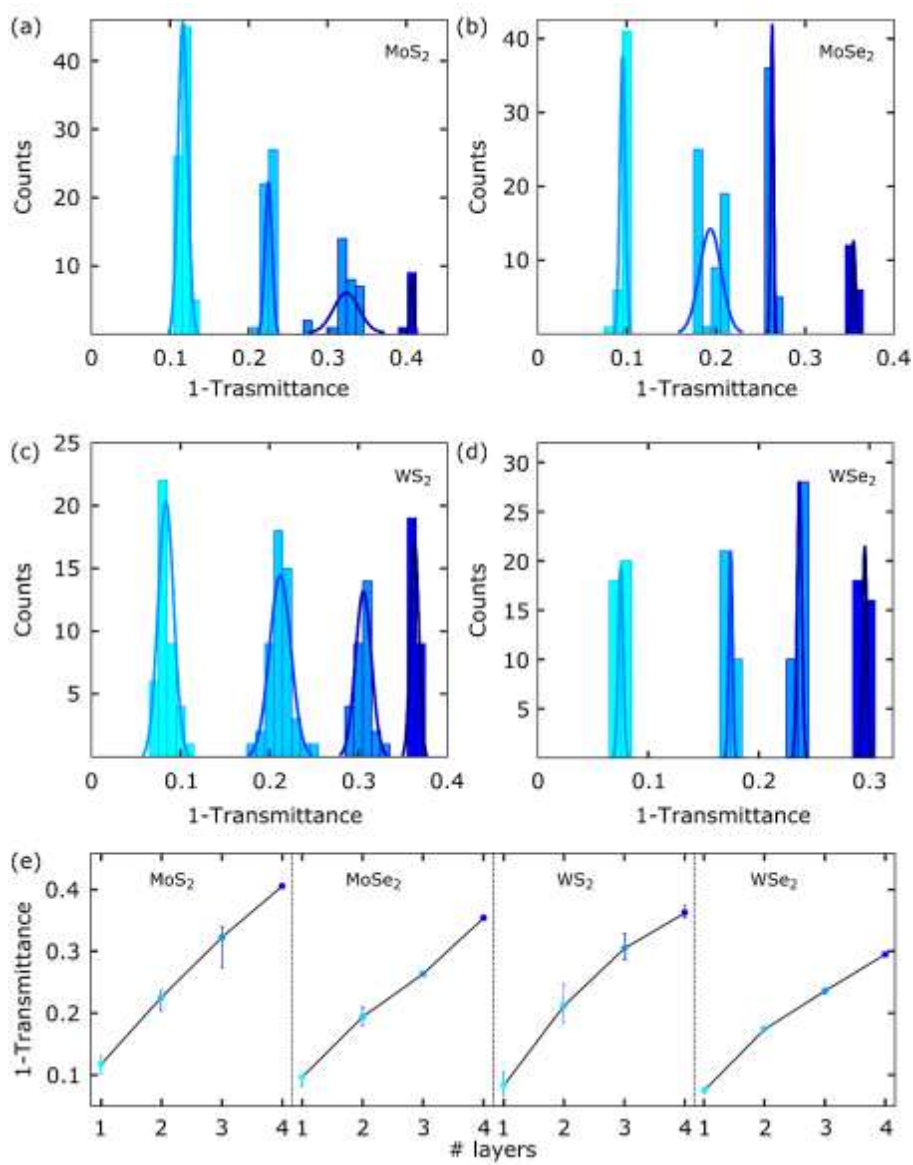

**Figure 4.** Histograms of the 1-transmittance value in the blue channel for (a) $MoS_2$, (b) $MoSe_2$, (c) $WS_2$ and (d) $WSe_2$ flakes. (e) Average of the 1-transmittance values for 1L to 4L in each material.

# Electronic Supplementary Material

# Thickness determination of $MoS_2$, $MoSe_2$, $WS_2$ and $WSe_2$ on transparent stamps used for deterministic transfer of 2D materials

*Najme S. Taghavi[1,2], Patricia Gant[1](✉), Peng Huang[1,3], Iris Niehues[4], Robert Schmidt[4], Steffen Michaelis de Vasconcellos[4], Rudolf Bratschitsch[4], Mar García-Hernández[1], Riccardo Frisenda[1](✉), Andres Castellanos-Gomez[1](✉)*

[1] Materials Science Factory, Instituto de Ciencia de Materiales de Madrid (ICMM), Consejo Superior de Investigaciones Científicas (CSIC), Sor Juana Inés de la Cruz 3, 28049 Madrid, Spain.

[2] Faculty of Physics, Khaje Nasir Toosi University of Technology (KNTU), Tehrān 19697 64499, Iran.

[3] State Key Laboratory of Tribology, Tsinghua University, Beijing 100084, China.

[4] Institute of Physics and Center for Nanotechnology, University of Münster, 48149 Münster, Germany.

————————————————

Address correspondence to Patricia Gant (patricia.gant@csic.es), Riccardo Frisenda (riccardo.frisenda@csic.es) and Andres Castellanos-Gomez (andres.castellanos@csic.es)

**INFORMATION ABOUT ELECTRONIC SUPPLEMENTARY MATERIAL**.
1. Constructing a thickness map from transmission mode optical microscopy image
2. Blue channel transmittance for flakes thicker than 4 layers

**Constructing a thickness map from transmission mode optical microscopy image**

Interestingly one can directly convert a transmission mode optical microscopy image into a thickness map by using this quantitative analysis of the blue channel. First, the blue channel of the transmission mode image is extracted and the average intensity of the substrate ($T_0$) is measured. Then, we calculate 1-$T/T_0$ to each pixel of the blue channel image, where T is the blue intensity at each pixel of the image. Under such transformation the substrate value becomes ~0 and the value on the flakes can be directly compared to the values in the histograms displayed in Figure 4. Indeed, by selecting a colormap accordingly to the 1-$T/T_0$ histograms in Figure 4 the map directly displays the number of layers.

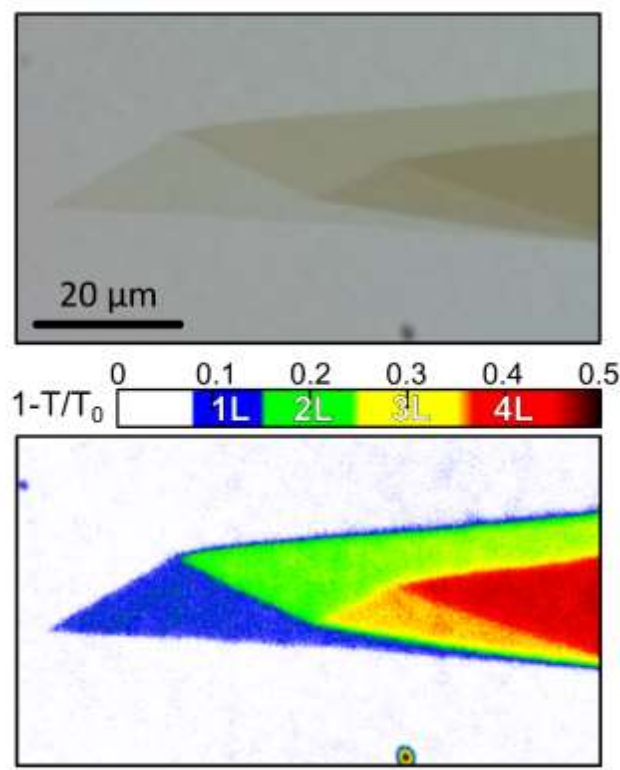

Figure S1. Transformation of a transmission mode optical microscopy image (top) into a thickness map (bottom) by applying the formula 1-$T/T_0$, where T is the intensity of each pixel and $T_0$ is the average intensity of the substrate. By selecting a colormap accordingly to the 1-$T/T_0$ histograms in Figure 4 the map displays the number of layers.

**Blue channel transmittance for flakes thicker than 4 layers**

In the main text we focused on the analysis of the transmittance of flakes 1L to 4L thick because we had plenty of statistics in that thickness range and because we used complementary techniques like Raman spectroscopy, photoluminescence and micro-reflectance/transmittance to double-check the assessed number of layers. The uncertainty of the number of layers assessment done by those techniques increases substantially for flakes thicker than 4-5 layers. Figure S2 shows the analysis of the transmittance of the blue channel for TMDC flakes thicker than 4 layers.

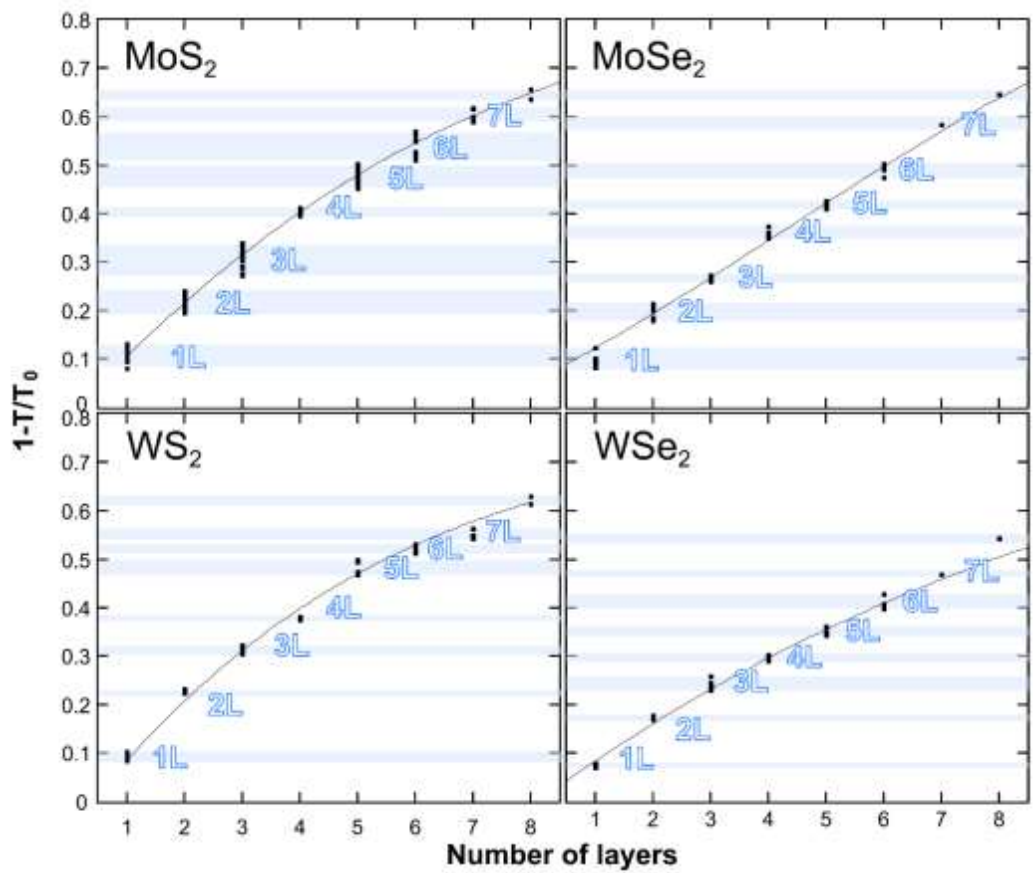

Figure S2. 1 - transmittance of the blue channel measured for $MoS_2$, $MoSe_2$, $WS_2$ and $WSe_2$ flakes with thicknesses in the range of 1L to 8L.

.